\documentstyle[preprint,aps,12pt]{revtex}

\tighten

\preprint{CU-TP-705, UTHEP-316, 1995}

\begin{document}

\draft

\title{What Thermodynamics tells about
 QCD Plasma \\ near Phase  Transition}

\author{M.\ Asakawa\footnote{e-mail: yuki@nt1.phys.columbia.edu}
and T.\ Hatsuda\footnote{e-mail: hatsuda@nucl.ph.tsukuba.ac.jp}
}

\address{Physics Department, Brookhaven National Laboratory, Upton,
NY 11973, USA}

\address{$^*$ Department of Physics, Columbia University,
 New York, NY 10027, USA}

\address{$^{\dagger}$ Institute of Physics, University of Tsukuba,
 Tsukuba, Ibaraki 305, Japan}

\date{August 24, 1995}

\maketitle

\begin{abstract}

 Due to a rapid change of the
 entropy density $s(T)$ across the critical temperature $T_c$ of the
 QCD phase transition, the pressure $P(T)$
 and the energy density $e(T)$ above $T_c$ generally
 deviate from their Stefan-Boltzmann values.
 We shall demonstrate this both analytically and numerically for
 a general class of $s(T)$ consistent with thermodynamical constraints
 and make a qualitative comparison of the result with the lattice QCD data.
 Quantities related to $ds(T)/dT$ such as the specific heat
 and sound velocity are also discussed.

\end{abstract}



\renewcommand{\thepage}{\arabic{page}}
\setcounter{page}{1}

\parskip=2pt

Revealing the precise nature of the quark-hadron
 phase transition is one of the central issues in recent
 lattice and analytical studies of hot QCD (see the reviews
 \cite{review}.)  Furthermore, the behavior of the bulk quantities such
 as the entropy density $s(T)$, energy density $e(T)$ and pressure $P(T)$
 as a function of temperature $T$ is relevant to the
 formation and evolution of the quark-gluon plasma in the
 ultrarelativistic heavy ion collisions planned at BNL and CERN.

Lattice QCD calculations have been providing interesting data
 on the plasma properties above the critical temperature $T_c$ \cite{review}.
  Among them are, however, some unexpected features:
(a) $P/T^4$ approaches the Stefan-Boltzmann limit very slowly
 as $T$ increases, (b)  $e/T^4$ has a peak just above $T_c$ and
approaches an asymptotic value from above as $T$ increase,
 and (c) $e-3P \neq 0 $ above $T_c$ i.e.
 a large deviation from the ideal gas behavior is seen.
 They are sometimes identified as indications  of the
 non-perturbative nature of the quark-gluon plasma.

In this letter, under generic assumptions on $s(T)$
 consistent with the thermodynamic inequality, we
 examine what is the expected behavior of $e(T)$, $P(T)$
 and other quantities such as the specific heat $C_V (T)$
 and the sound velocity $c_s(T)$. We show that
 most of the ``unexpected'' behaviors
 measured in the lattice QCD simulations \cite{review,blum,lae}
 can be explained at least qualitatively as
 a simple consequence of the rapid increase of $s(T)$ around
 $T_c$.

Our basic observation is that all the above quantities are simply
 parametrized by the entropy density $s(T)$ which
 depends only on the active degrees of freedom
 and is free from the complexities of the vacuum structure of QCD.
 This is seen by the thermodynamic relations at zero chemical
 potential such as
\begin{eqnarray}
P(T) & = & \int_0^T s(t) dt,
\label{pressure} \\
e(T) & = & Ts(T) - P(T),
\label{energy}
\end{eqnarray}
and
\begin{eqnarray}
C_V & = & T { \partial s(T) \over \partial T},
\label{cv} \\
c_s^2 & = & {\partial P \over \partial e} = \left[
 \partial \ln s(T) \over \partial \ln T \right]^{-1} .
\label{cs}
\end{eqnarray}
In eq.(\ref{pressure}), $P(T)$ is normalized by $P(T=0)=0$,
 which also leads to $e(T=0)=0$ from eq.(\ref{energy}).
 The same normalization is adopted also in lattice calculations.

Since we are interested only in the gross behavior of the
 thermodynamic quantities and not in the precise order
 of the phase transition \cite{order}, we make a smooth
 interpolation of $s(T)$  between the
 the hadronic gas at low $T$
 and the quark-gluon plasma at high $T$.
 Note, however, that one cannot make arbitrary
 parametrization of $s(T)$ since it is constrained by
 the thermodynamic inequality and the Nernst's theorem \cite{landau}
\begin{eqnarray}
{\partial s(T) \over \partial T} > 0, \ \ \ \ s(0)=0.
\label{ine}
\end{eqnarray}

A simplest possible parametrization
satisfying eq.(\ref{ine}) reads
\begin{eqnarray}
s(T) = s_h(T)w_h(T) + s_q(T)w_q(T),
\label{entropy}
\end{eqnarray}
where $w_h(T) = 1 - w_q(T) $ and $w_q(T)$ is an increasing function.
In this letter,
we take the following form for $w_q(T)$, but our conclusion
is not limited to this specific choice of $w_{h,q}(T)$:
\begin{eqnarray}
w_q(T)=
{n \left( 1 + \tanh \left( {T-T_c \over \Gamma} \right) \right)
 \over
 m \left( 1 - \tanh \left( {T-T_c \over \Gamma} \right) \right)
 + n \left( 1 + \tanh \left( {T-T_c \over \Gamma} \right) \right)}.
\label{entropy1}
\end{eqnarray}
Here $s_h(T) \equiv 12(\pi^2/90)T^3$ and
 $s_q(T) \equiv 148(\pi^2/90)T^3$ are the
 entropy densities of massless free gas with two flavors in the
 hadronic phase (pion gas) and the quark-gluon phase,
 respectively.  The interaction between particles and the
 quark masses are neglected just for simplicity.
  $2\Gamma$ sets
 the width of the transition region.  For $T$ satisfying
 $|T-T_c| \gg \Gamma$, $s(T)$ approaches
 asymptotically to $s_h(T)$ below $T_c$ and to $s_q(T)$
 above $T_c$.  $m$ and $n$ are introduced to consider the asymmetric
 superposition of the two phases around $T_c$, but we shall
 take $m=n$ in the following.

The equation of state obtained from eq.(\ref{entropy})
 is more general than that of the bag model \cite{satz}
 in the sense that (a) we need not refer to phenomenological parameters
 such as the bag constant, and (b) we can treat not only the strong first order
 phase transition but also the crossover by changing $\Gamma$ in a
thermodynamically consistent way.  When $\Gamma=0$, our equation of state
 is equivalent to the bag model one with a bag constant
 $4B=(s_q(T)/T^3 - s_h(T)/T^3)T_c^4$.

In Fig.1, $s(T)$ with $m=n=1$ is shown for
 $\Gamma/T_c = 0, \ 0.05$ and 0.25.
 Lattice measurements show a rapid variation
 of $s(T)$ in a narrow region of $T$
 (with a width $\sim 10$ MeV)\cite{christ}, which is
 similar to Fig.1 with $\Gamma /T_c = 0.05$.
 Inclusion of the perturbative $\alpha_s$
 corrections above $T_c$ and the interactions of hadrons
 below $T_c$ modifies the absolute value of $s(T)$
 as well as its $T$ dependence, but it does not
 change our conclusions qualitatively.

Once $s(T)$ is given, it is straightforward to calculate
 other quantities from
 eqs.(\ref{pressure},\ref{energy},\ref{cv},\ref{cs}).
 In Fig.2, $e/T^4$, $P/T^4$ and $(e-3P)/T^4$
 are shown as a function of temperature for $\Gamma/T_c = 0.05$.
 They can be directly compared with the lattice results
 \cite{review,blum,lae} at least qualitatively. In fact, they
 look quite similar:

(i) $P/T^4$ increases rather slowly above $T_c$
 both on the lattice \cite{blum,lae} and in Fig.2.
 Since $P(T)$ is given as an integral of $s(T)$, it
 is quite natural to expect such a continuous and slow rise.
 When $\Gamma =0$, $P(T>T_c)$ has an analytic  form
\begin{eqnarray}
P(T)/P_{SB}(T) \sim 1 - (T_c/T)^4,
\label{ppsb}
\end{eqnarray}
where $P_{SB}(T)$ denotes the Stefan-Boltzmann value.
 Eq.(\ref{ppsb})  tells us that
 $P(T)/P_{SB}(T) = 50 \% \ (90\% )$ for
 $T/T_c = 1.2 \ (1.8)$ independent of the details of the
 dynamics.  If $\Gamma$ is increased, $P/T^4$
 has even weaker $T$ dependence.
 Thus the major part of the deviation of $P(T)$
 from $P_{SB}(T)$ can be accounted for
 without introducing non-perturbative interactions
 of quarks and gluons above $T_c$.

(ii) $e/T^4$ has a peak just above $T_c$ in Fig. 2
(note that $e(T)$ itself is
 a monotonically increasing function of $T$),
  which  is also seen in
 lattice simulations \cite{blum}.
 In our case, this peak is a simple consequence of the rapid
 increase of $s(T)$ and the slow rise of $P(T)$ above $T_c$
 (see the definition eq.(\ref{energy})).
 The width of the peak is correlated with the
 slow rise of $P/T^4$.
 One can even prove analytically that there must exist
 a peak around $T_c$ from  the following
 relation satisfied by  arbitrary weight
functions $w_{h,q}(T)$,
\begin{eqnarray}
T^5 \left (\frac{e}{T^4} \right )'
& = & T^2(s_h(T)w_h'(T)+s_q(T)w_q'(T)) \nonumber \\
  & & - \int_0^T \ t\ (s_h(t)w_h'(t)+s_q(t)w_q'(t)) dt,
\label{del}
\end{eqnarray}
where the prime denotes a derivative with respect to $T$
 and we have used a fact $s_{h,q}(T)$=(constant)$\times T^3$.
 Under the conditions that
 $w_h(t) + w_q(t) =1$, $0 \le w_q(0) < 1$,
 $w_q'(t) > 0$ and $w_q'(t \rightarrow \infty ) = o(1/t^4)$,
 $(e/T^4)'$ is positive (negative) for low (high)
 $T$ and has a zero near $T_c$.
 These conditions are all satisfied in our case, and as a result
 $e/T^4$ has a peak.
 One should also note that, for high enough $T$,
 $(e/T^4)' \sim - \Delta(T)/T < 0$ with
 $\Delta \equiv (e-3P)/T^4$.

(iii) $\Delta(T) = (e-3P)/T^4$ is so
 called the ``interaction measure'' and has a peak
 near $T_c$ both on the lattice \cite{blum,lae}
 and in Fig.2. Again, the rapid increase of
$s(T)$ (i.e. the liberation of quarks and gluons)
 is the reason for this peak, which is also seen from the general formula
\begin{eqnarray}
\Delta (T) = {1 \over T^4}
 \int_0^T \ t\ (s_h(t)w_h'(t)+s_q(t)w_q'(t)) dt.
\label{delta}
\end{eqnarray}
$\Delta$ vanishes for $T=0$ and $ \infty$ and has a peak around $T_c$
 independent of the details of the dynamics.
 Also the peak becomes sharper and
 its height becomes higher as $\Gamma$ decreases.
 When $\Gamma=0$, one can rewrite $\Delta$ as
 $\Delta (T>T_c) = (s_q/T^3 - s_h/T^3)(T_c/T)^4$ which
 is equivalent to the bag model  formula without $\alpha_s$
 corrections $\Delta(T>T_c) = 4B/T^4$ \cite{satz}.
 Note here that, if we normalize the pressure by $P(\infty) = P_{SB}$
  instead of $P(0)=0$, the peak of $\Delta(T)$ does not arise
  and the naive Stefan-Boltzmann law is realized above $T_c$.
  In this sense,  the $1/T^4$ tail of $\Delta$ at high $T$ in Fig.2
 can be interpreted as an artifact of the normalization and
 has nothing to do with non-perturbative interactions
 of quarks and gluons in the quark-gluon plasma.

 In Fig.3,  $c_s^2$ is shown
 for $\Gamma/T_c=0.05$.
   Since the equation of state becomes soft near the critical
 region, the sound velocity slows down. This is the reason why
 $c_s^2$ has a sudden drop in the narrow region
 $T_c - \Gamma < T < T_c + \Gamma$. This sudden change
  is in contrast to the broad peak or slow rise of the quantities
 in Fig.2. The difference comes from the fact that
  $c_s^2$ is related to the derivative of $s(T)$ while
 $e(T)$ and $P(T)$ are related to the integral of $s(T)$.
  To see the effect of the finite pion mass on $c_s^2$ below $T_c$,
 $c_s^2$ using $s_h(T)$ with $m_{\pi}=140 $MeV (and $T_c=180$MeV) is shown
 in Fig.3 by the dashed line.  The effect of finite
 $m_{\pi}$ on the other quantities is small since
 they are small in any way at low $T$.
 The heat capacity $C_V$ has also a sharp peak at $T_c$,
 since one needs to supply a large amount of  heat to increase $T$
 across $T_c$ to liberate quark-gluon degrees of freedom.

Although we need not refer to the vacuum parameters such as the bag constant
  in our approach, it might be instructive to
 introduce an effective bag constant above $T_c$ defined as
 a deviation of the pressure from its Stefan-Boltzmann value,
\begin{eqnarray}
B(T> T_c) = -(P(T) - P_{SB}(T))/T^4,
\end{eqnarray}
where we have neglected all the $\alpha_s$ corrections.
 $B(T)/T_c^4$ is shown in Fig.4. The asymptotic value $B( \infty)$
 is the one usually used in the bag model.
 In our case, $B(\infty)$ depends on how one parametrizes $s(T)$:
 as $\Gamma$ is increased, $B(\infty)$ also increases.

 The main conclusion of this letter is that,
 whenever the entropy density has a rapid change near $T_c$,
 $e/T^4$, $P/T^4$ and $\Delta$ behave as
 we know from the lattice simulations.
  Furthermore, we need not refer to the vacuum condensate
 or the bag constant to see this fact.
 We believe that our approach based on the parametrization of $s(T)$
 provides a transparent and thermodynamically consistent
 way to study the qualitative feature of the bulk plasma quantities.
 Although we have taken a simplest possible form of
 $s(T)$, inclusion of the quark masses,  hadronic interactions,
 $\alpha_s$ corrections
 and chemical potentials is straightforward.
 Possible non-perturbative effects at $T_c < T < 3 T_c$
   \cite{lae,RG}
 appear  as a deviation from the basic curves
 in Fig.2 and could be included
  by adjusting the functional form of  $s(T)$.
 However, it is not an easy task to identify
 true non-perturbative effects from the lattice data,
 since Fig.2 has already similar behavior with the lattice
 data, and   furthermore  the finite volume effect and perturbative
 corrections are  still large in the
 current lattice simulations \cite{review}.

\acknowledgements

We thank Teiji Kunihiro for useful comments and discussions, and
 Rob Pisarski for warm hospitality at BNL where
 most part of
 this work was completed.
 We also thank O. Miyamura  and M. Gyulassy
 for private communications who have already realized
 a part of the present results in the context of the bag model.
 M. A. was supported by the Director, Office
 of Energy Research, Division of Nuclear Physics of the Office of High Energy
 and Nuclear Physics of the U.S. Department of Energy under Contract
 No.DE-FG02-93ER40764.
 T. H. was supported in part by the Grants-in-Aid of the
 Japanese  Ministry of  Education (No. 06102004).


\vspace{3cm}

\centerline{Figure Captions}

\noindent
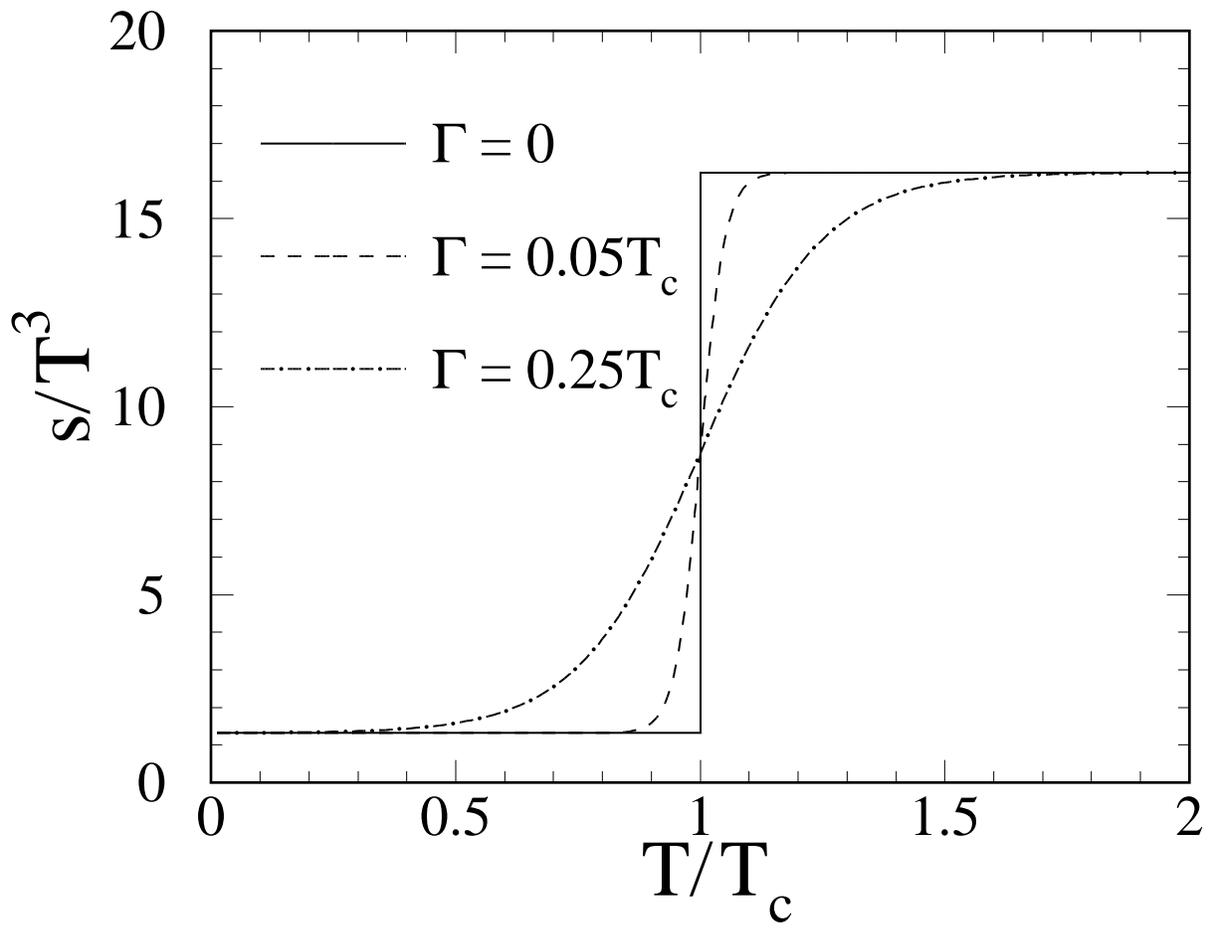
Fig.1: $s/T^4$ as a function of $T$ for three
 different values of the width parameter $\Gamma$.

\noindent
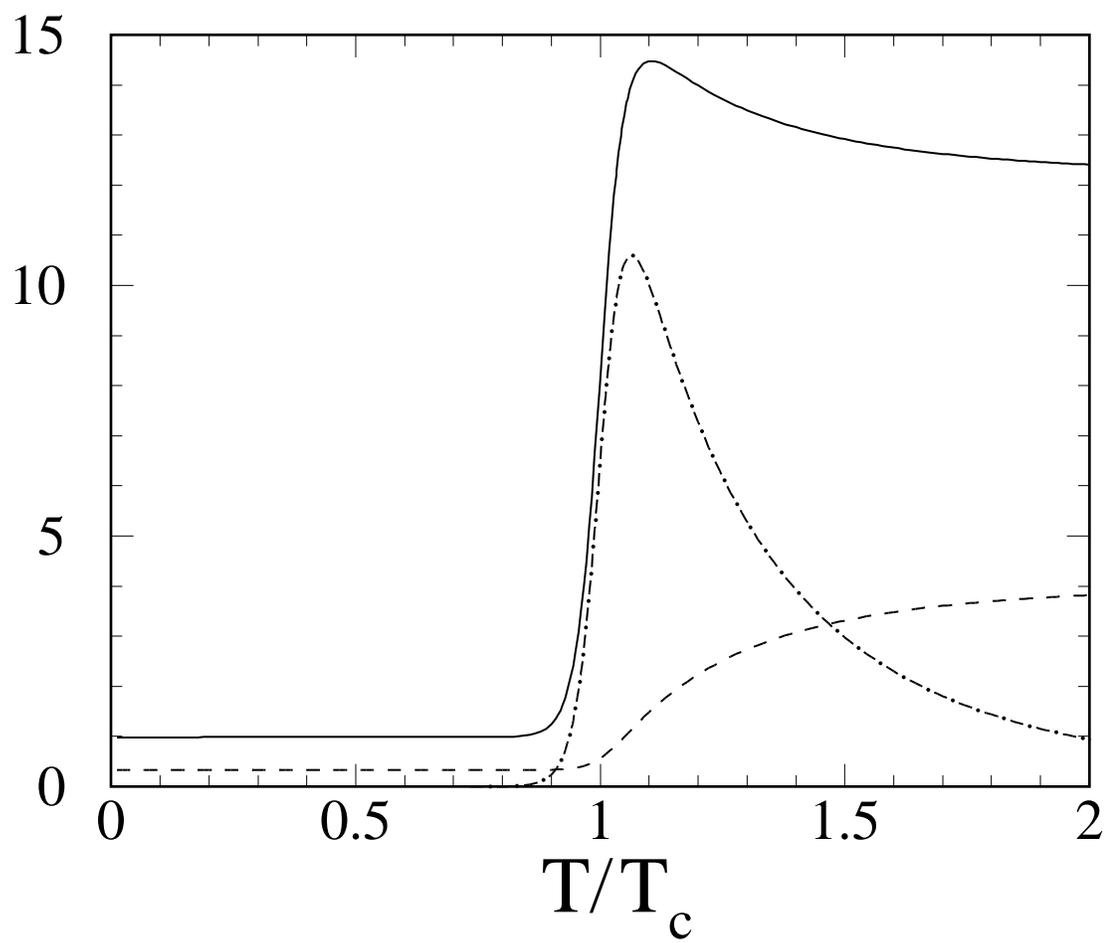
Fig.2: The solid line, the dashed line and the dash-dotted
 line correspond to $e/T^4$, $P/T^4$ and $\Delta$, respectively.
 $\Gamma/T_c=0.05$ is chosen.

\noindent
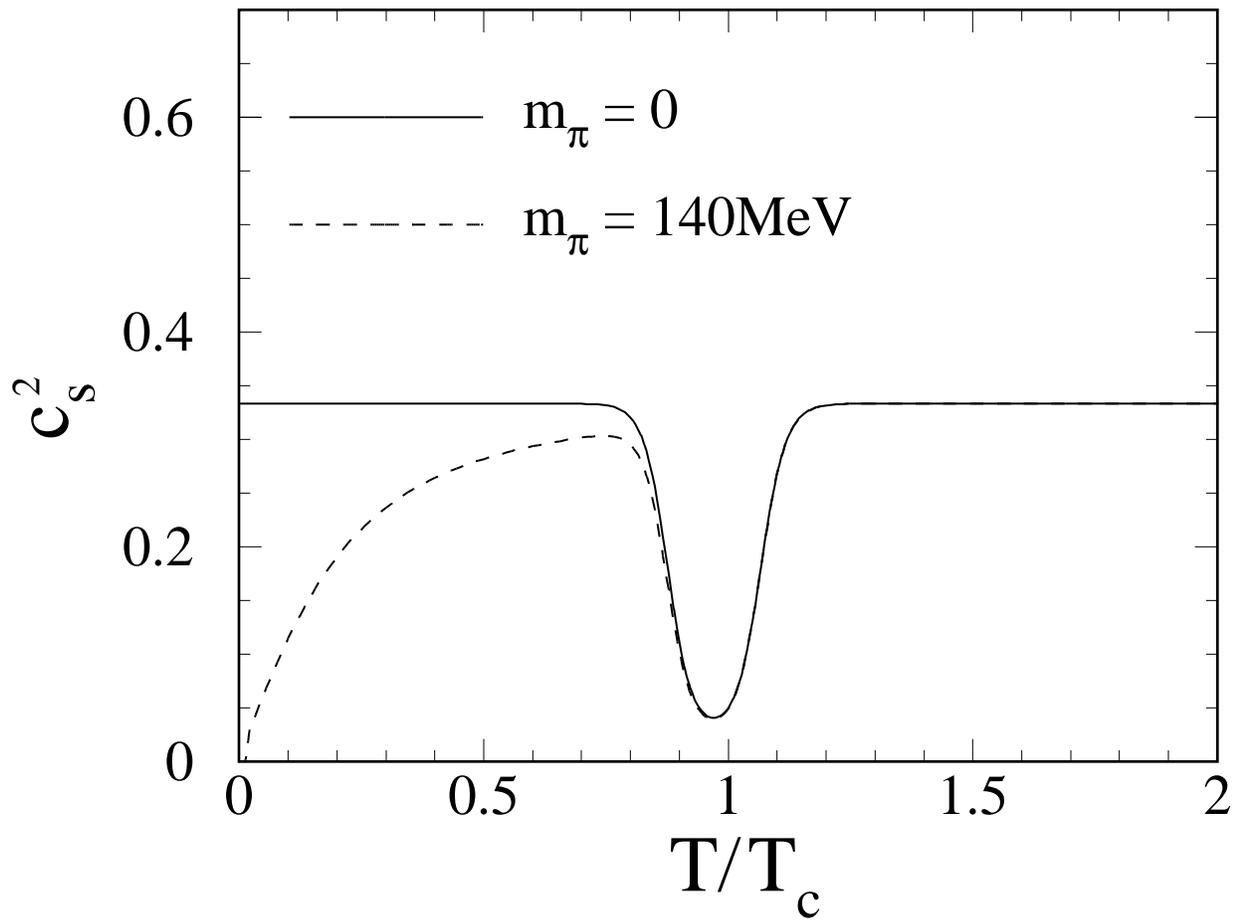
Fig.3: Squared sound velocity $c_s^2$ as a function of $T$.
 Solid (dashed) line corresponds to $m_{\pi}=0 \ (140)$ MeV.

\noindent
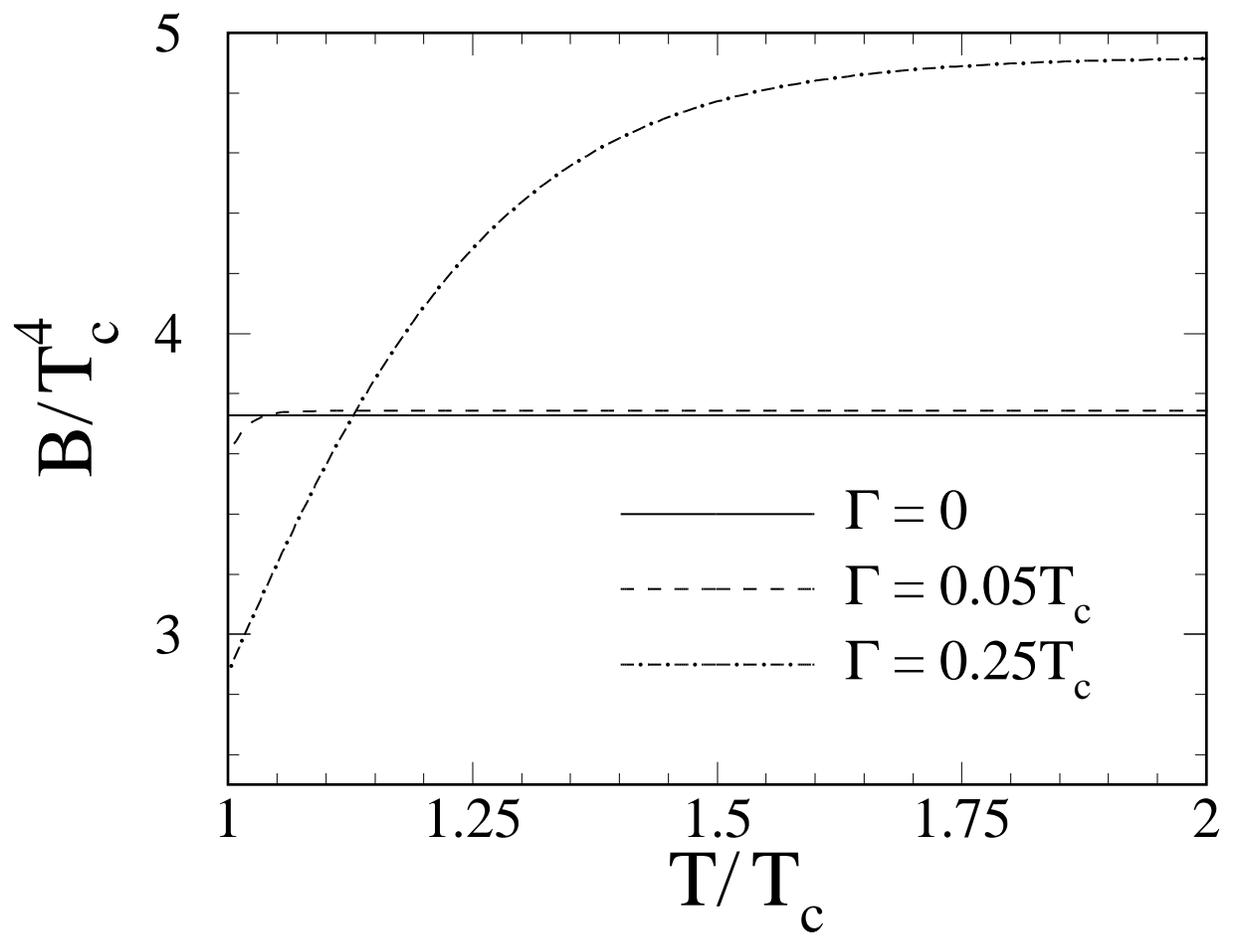
Fig.4:  ``Effective'' bag constant $B(T)/T_c^4$ for $T > T_c$.
 Solid, dashed and dash-dotted lines correspond to
 $\Gamma/T_c = 0,\ 0.05$ and 0.25, respectively.

\end{document}